\documentclass[pdflatex,sn-mathphys-num]{sn-jnl}% Math and Physical Sciences Numbered Reference Style
\usepackage{graphicx}%
\usepackage{multirow}%
\usepackage{amsmath,amssymb,amsfonts,braket,dsfont}%
\usepackage{amsthm}%
\usepackage{mathrsfs}%
\usepackage[title]{appendix}%
\usepackage{xcolor}%
\usepackage{textcomp}%
\usepackage{manyfoot}%
\usepackage{booktabs}%
\usepackage{algorithm}%
\usepackage{algorithmicx}%
\usepackage{algpseudocode}%
\usepackage{listings}%
\usepackage{quantikz}
\usepackage{enumitem}

\theoremstyle{thmstyleone}%
\theoremstyle{thmstyletwo}%

\theoremstyle{thmstylethree}%

\begin{document}

\title[Article Title]{Quantum Simulation of Collective Neutrino Oscillations in Dense Neutrino Environment}


\author*[1,2]{\fnm{Shvetaank} \sur{Tripathi}}\email{shvetaank05@gmail.com}

\author*[1]{\fnm{Sandeep} \sur{Joshi}}\email{sjoshi@barc.gov.in}

\author*[1]{\fnm{Garima} \sur{Rajpoot}}\email{garimar@barc.gov.in}

\author*[1,2]{\fnm{Prashant} \sur{Shukla}}\email{pshukla@barc.gov.in}

\affil[1]{\orgdiv{Nuclear Physics Division}, \orgname{Bhabha Atomic Research Institute}, \orgaddress{\street{Trombay}, \city{Mumbai}, \postcode{400085}, \state{Maharashtra}, \country{India}}}

\affil[2]{\orgname{Homi Bhabha National Institute}, \orgaddress{\street{Anushaktinagar}, \city{Mumbai}, \postcode{400094}, \state{Maharashtra}, \country{India}}}

\abstract{Inside dense neutrino gases, such as neutron star mergers or core-
collapse supernovae, collective neutrino effects cause the transformation of one 
neutrino flavour into another. Due to strong neutrino self-interactions in these environments, there is prevalence of flavour swapping. Considering these environments to be isotropic and homogeneous, we present a study of collective neutrino oscillations by simulating such a system on a noisy quantum simulator 
(Qiskit \texttt{AerSimulator}) and a quantum processor (\texttt{ibm\_brisbane}).
We model the effective Hamiltonian governing neutrino interactions and by applying the Trotter–Suzuki approximation, decompose it into a tractable form suitable for quantum circuit implementation of the time-evolution propagator. Encoding the neutrino state for a system of two- and three-neutrinos onto qubits, we compute the time evolution of the inversion probability relative to the initial product state. Furthermore, we present quantum circuits to evaluate the concurrence as a measure of entanglement between the neutrinos.}


\maketitle

\section{Introduction}\label{sec:Intro}
Neutrino oscillation refers to the quantum mechanical phenomenon in which a neutrino
created with a specific flavour (electron, muon, or tau) can later be detected as a 
different flavour. This occurs due to the fact that the flavour eigenstates are non-
trivial superpositions of the neutrino mass eigenstates, combined with the non-
degenerate masses of these mass eigenstates~\cite{Pontecorvo:1957cp,Maki:1962mu}. 
As neutrinos propagate, the relative phase between their mass components evolves with 
time, leading to periodic flavour transitions. This is a manifestation of quantum 
interference and is observed in both vacuum and 
matter~\cite{SajjadAthar:2021prg,Giunti:2007ry,Pontecorvo:1967fh,Kuo:1989qe}. 
The presence of matter modifies the oscillation parameters through coherent forward 
scattering, a phenomenon known as the Mikheyev–Smirnov–Wolfenstein (MSW) 
effect~\cite{Wolfenstein:1977ue,Mikheyev:1985zog}, which plays a crucial role in 
neutrino propagation in astrophysical environments such as the Sun and neutron star.
Experimental confirmation of neutrino oscillations has provided direct 
evidence for non-zero neutrino masses, requiring an extension of the Standard Model 
of particle physics~\cite{Super-Kamiokande:1998kpq,SNO:2002tuh,KamLAND:2002uet}.

In a medium with high density of neutrinos, such as
core collapse supernovae, the self-interaction
of neutrinos causes the flavour swapping, which is also called collective
neutrino oscillations~\cite{Duan:2010bg, Volpe:2023met}.
Such environments are generally inhomogeneous and anisotropic, 
making the modelling of neutrino dynamics highly complex. 
Several theoretical studies suggest the emergence of entanglement during
the collective neutrino flavour 
oscillations~\cite{Pantaleone:1992xh,Sigl:1993ctk,Pastor:2001iu,
Raffelt:2002tu,Raffelt:2007yz,Blasone:2007vw,Duan:2010bg}.
Roggero $\emph{et\ al.}$~\cite{Roggero:2021asb} demonstrate that 
in collective neutrino systems many-body correlations crucially impact 
dynamical observables. They compute survival probabilities and half-chain 
entanglement entropy in a spin-model framework, revealing logarithmic 
scaling of entanglement with system size.
The entanglement during the flavour evolution of the states is also
found to have some correlation with spectral spliting of neutrino 
energy~\cite{Siwach:2024jet} and has dependence upon different
mass-ordering~\cite{Siwach:2022xhx}.
The time-evolution equations for self-interacting neutrinos are non-linear,
coupled differential equations that scale poorly with system size and require 
significant computational resources for working with many-neutrino system.
Ref.~\cite{Roggero:2022hpy} emphasizes that classical approaches, such as
mean-field or semi-classical approximations, are fundamentally limited in 
their ability to capture quantum correlations.
In particular, they fail to account for the growth of entanglement entropy, 
which is a hallmark of exact many-body quantum evolution. 
As a result, classical simulations underestimate decoherence, overlook quantum
interference effects, and cannot accurately describe the collective flavour
oscillation patterns that emerge in a fully quantum treatment. 

In recent times, advancements in the field of
quantum computing has sparked an increased interest in simulating 
phenomena such as neutrino oscillations on a quantum 
computer~\cite{DiMeglio:2023nsa, Bauer:2022hpo}.
Encoding the neutrino flavour as the state of a qubit, the time evolution of 
neutrinos can be simulated on a quantum processor. 
Simulations of coherent time evolution of two- and three-flavour 
neutrino oscillations, both in vacuum and in matter provide a fundamental
and robust foundation for further investigation.
These simulations have been performed on multiple physical hardware, such as 
superconducting qubits \cite{Arguelles:2019phs, Nguyen:2022snr, Joshi:2025jxe},
trapped ions \cite{noh2012quantum} and nuclear magnetic resonance (NMR) processor 
\cite{Singh:2024vpu}.

To simulate the collective neutrino oscillations, different approaches 
were taken which includes the calculation of energy eigenvalues
either by the diagonalisation~\cite{Patwardhan:2019zta} or by the
decomposition of total Hamiltonian into smaller H-blocks~\cite
{Yeter-Aydeniz:2021olz}, which also helps in noise reduction.
Turro et al.~\cite{Turro:2024shh} introduce both qutrit and qubit encodings
for simulating three-flavour collective neutrino oscillations, overcoming
limitations of the traditional two-flavour approximation.
They design optimized quantum circuits to capture three-flavour dynamics
and demonstrate their feasibility by running simulations on IBM (qubits)
and Quantinuum (qutrits) devices, for systems of 2, 4, and 8 neutrinos.
Spagnoli et al.~\cite{Spagnoli:2025etu} presents the qubit and qutrit 
encodings for the full three-flavour neutrino system, paying special 
attention to Trotterization errors. 
They successfully execute time-evolution circuits on superconducting 
hardware—IBM’s Torino (qubits) and AQT (qutrits)—for a two-neutrino system.
Applying robust error mitigation techniques, they achieve results consistent 
with ideal simulations, noting the qutrit-based circuits avoid probability 
leakage issues common in qubit mappings.
These results open a whole new area to study the complex 
astrophysical systems having an all-to-all interaction.
While such simulations can provide a very good picture of the system 
involved, the errors caused during the evolution of coherent states 
are also inevitable and they increases with the
size of the system. 
Thus, study of such errors and noises is equally important, as 
being done in Refs.~\cite{Amitrano:2022yyn,Siwach:2023wzy}. 

Occurence of entanglement during collective neutrino oscillations is
due to the self-interactions of neutrinos in dense environment.
Here, the neutrinos do not evolve independently, which could be the
case in vacuum and matter oscillations but coherently interact with one 
another, which results in flavour swapping. 
Due to this coupling, the total wavefunction of the system cannot be 
written as a product of individual neutrino wavefuntions. 
In quantum computation, entanglement quantifies quantum correlations 
using measures like concurrence, tangle, and entanglement
entropy~\cite{Wootters:1997id,Hill:1997pfa,Wootters:2001ceg,Swain:2017jso}. 
These metrics capture non-classical features of quantum systems 
and validate the use of quantum processors for simulating systems 
with complex, all-to-all interactions. 
By indicating the degree of interaction between neutrinos, 
entanglement analysis deepens our understanding of system dynamics.
In this regard, the study of neutrino flavour oscillations
to show the evolution of bi- and tri-partite  entanglement between 
the coherent states of neutrinos has been done in 
Refs.~\cite{Jha:2020qea,KumarJha:2020pke,Jha:2021itm}.
A key contribution in this context is Ref.~\cite{Hall:2021rbv}, 
which investigates the dynamics of entanglement in a four-neutrino
system using both entanglement entropy and pairwise concurrence 
as quantitative measures. 
This study demonstrates how entanglement emerges and evolves in 
neutrino systems undergoing coherent forward scattering, offering 
a concrete framework to explore many-body quantum correlations in 
collective neutrino oscillations. 
It also emphasizes the challenges posed by noise in 
current quantum hardware and the critical role of error mitigation 
techniques in reliably extracting entanglement properties. 
These insights are foundational for validating quantum simulation 
approaches in neutrino physics.
However, the quantum circuit implementation of collective neutrino 
effects was not detailed in that work, leaving room for further studies 
to explore circuit-level realizations of such many-body dynamics.

Employing quantum simulation techniques, we model two- and three-
neutrino systems to study the time evolution of the inversion 
probability for neutrino flavour product state and the pairwise concurrence
between individual neutrino flavours.
This manuscript highlights the effects of noise 
and complexity of the circuit on the evolution of these states.
Here we provide the detailed quantum circuits required to simulate
the system of neutrinos in a supernova.
While this work presents the basic concept of neutrino interactions 
and evolution on a small-scale quantum system, scaling to larger systems 
introduces significant challenges, including the exponential growth of 
Hilbert space, increased circuit depth, and greater sensitivity to noise, 
all of which require advanced error mitigation and more powerful
quantum hardware.
Scaling such many-body treatment of the system on a quantum processor to a 
large number of neutrinos will facilitate the development of theoretical aspects 
of neutrino physics in dense neutrino gases.

The structure of this paper is as follows:
Section~\ref{sec:HamiltonianTimeEvol} outlines the construction of
the Hamiltonian and time evolution of the states, inspired by the supernova 
neutrino bulb model.
Section~\ref{sec:QuantSim2Neut} details the quantum simulation and calculation
of state inversion probability for two- and three- neutrino system. 
Entanglement measurements for two-neutrino system are discussed in 
Section~\ref{sec:Entang}. 
Section~\ref{sec:Summ} summarizes the findings and 
concludes the study.

\section{Hamiltonian and effective unitary of propagation}\label{sec:HamiltonianTimeEvol}
In dense neutrino gases like supernovae, the initial fluxes of $\nu_\mu$,
$\bar{\nu}_\mu$, $\nu_\tau$ and $\bar{\nu}_\tau$ are almost the same 
(See Ch.10 of~\cite{barger2012physics}).
Thus, we will work in the limit of two-flavour neutrino oscillations.
The light flavour of a neutrino corresponds to the electron-type, 
while the heavy flavour can be either the $\mu-$ or $\tau-$type. 
The total Hamiltonian includes contributions
from vacuum oscillations, interaction with background matter, and neutrino self interaction~\cite{Bell:2003mg,Friedland:2003eh,Sawyer:2005jk,
Friedland:2006ke,Balantekin:2006tg}.
\begin{equation} \label{eq:netHint}
  H = H_{vac} + H_{e} + H_{\nu \nu},
\end{equation}
where, $H_{vac}$  is the vacuum Hamiltonian which is responsible 
for the flavour mixing in vacuum over astronomical distances.
$H_{e}$ represents the forward $\nu_{e}-e$ scattering inside matter
via $W$ exchange~\cite{Kuo:1989qe,Wolfenstein:1977ue,Barger:1980tf}.
$H_{\nu \nu}$ represents the neutrino-neutrino forward scattering via
$Z$ exchange ~\cite{Fuller:1987gzx,Notzold:1987ik}.
The coherent evolution of neutrino flavour state is governed by the 
Schr\"odinger's equation.:
\begin{equation}\label{eq:timeevol}
  i \frac{\partial \psi_\nu}{\partial t} = H \psi_\nu.
\end{equation}
We may write the vacuum and matter mixing Hamiltonian from Eq.~\ref{eq:netHint}
in terms of Pauli operators ($\sigma^x, \sigma^y, \sigma^z$) as : 
\begin{equation}\label{eq:H1}
  H_{vac} + H_e = \frac{1}{2}\sum_{i=1}^N\Big[\frac{\Delta m^2}{2E_i}\Big(-
  \cos2\theta_\nu \sigma_i^z +
  \sin2\theta_\nu \sigma_i^x\Big) + V_{CC} \sigma_i^z \Big], 
\end{equation}
where $E_i$ denotes the energy of neutrino, $\Delta m^2$ is the mass-squared
difference ($\Delta m^2= m_2^2- m_1^2$) and $\theta_\nu$ is the
vacuum mixing angle. 
$V_{CC}$ represents the charged-current potential, given by 
$V_{CC} = \sqrt{2}G_Fn_e$,
where $G_F$ is the Fermi constant and $n_e$ is the electron
density.
The neutrino-neutrino interaction Hamiltonian is given by:
\begin{equation}\label{eq:H2}
  H_{\nu\nu} = \sum_{i<j}^N\eta \Big(1 - \hat{q_i}\cdot\hat{q_j}\Big)
  \vec{\sigma_i}\cdot\vec{\sigma_j},
\end{equation}
where $\eta= G_Fn_\nu/(\sqrt{2}N)$ is the coupling strength with $N$
being the total number of neutrinos in consideration and $n_\nu$ being
the neutrino density. 
To obtain a simplified form of the Hamiltonian \eqref{eq:H1}, we make the follwing assumptions :
\begin{enumerate}[label=(\alph*)]
    \item In the region around  $\sim 100$ km from the core of the supernova, the neutrino density is much larger compared to the electron density. Thus, in this region, coherent vacuum oscillations dominate and we can assume $V_{CC}\rightarrow0$. 
    \item We use the neutrino bulb model~\cite{Duan:2006an}, 
    that assumes the spherical symmetry and 
    isotropic emission of the neutrinos from the core-collapse supernova. In this model the average coupling is obtained by averaging over the azimuthal angle of neutrino emission: $\textlangle 1 - \hat{q_i}\cdot\hat{q_j} 
    \textrangle = 1 - \cos\theta_i\cos\theta_j$. 
    \item Further we apply single angle approximation in which an average coupling between all the pairs of neutrinos is assumed. 
    This reduces the above expression, $1 - \cos\theta_i\cos\theta_j$ to 
    $1 - \cos\theta^{ij}$, with $\theta^{ij}$ being
    the angle between the momentum of two neutrinos.
    \item We assume all the neutrinos to be of same energy. 
    We take these energies to be, $E_\nu = \Delta m^2/4\eta$.  
\end{enumerate}
Thus, the total Hamiltonian for a system of $N$ self-interacting neutrinos, 
expressed in terms of the parameter $\eta$, can be written as~\cite{Hall:2021rbv}:
\begin{equation}\label{eq:netH}
  H = \sum_{k=1}^N \vec{b}\cdot\vec{\sigma_k} +
  \sum_{p<q}^N J^{pq}\vec{\sigma_p}\cdot\vec{\sigma_q}, 
\end{equation}
where $\vec{b} = (\sin2\theta_\nu, 0, -\cos2\theta_\nu)$ is an external field vector
determined by the vacuum mixing angle $\theta_\nu$, and $J^{pq} = 1 - \cos\theta^{pq}$
is the coupling strength between the $p^{\text{th}}$ and $q^{\text{th}}$ neutrinos.
The first term represents vacuum oscillations of individual neutrinos, while the second 
term encodes the flavour-changing interactions between neutrino pairs.

To simplify numerical implementation and to treat interactions in a pairwise fashion, we 
rewrite the Hamiltonian in Eq.~\ref{eq:netH} by grouping terms involving distinct neutrino 
pairs. This results in a reformulated Hamiltonian:
\begin{equation}\label{eq:reducedH}
  H = \sum_{p<q}^N \left[\frac{\vec{b}\cdot(\vec{\sigma_p}+\vec{\sigma_q})}{N - 1}
  + J^{pq}\vec{\sigma_p}\cdot\vec{\sigma_q}\right]
  = \sum_{p<q}^N \left( H_1^{pq} + H_2^{pq} \right),
\end{equation}
where $H_{1}^{pq}$ and $H_{2}^{pq}$ denote the single-body and interaction contributions
for each pair $(p,q)$ respectively. This decomposition is particularly useful for quantum
simulation, as it allows the total Hamiltonian to be expressed as a sum of smaller, two-
body Hamiltonians acting on qubit pairs, which we will see in later sections.
To study the time evolution of the system, we use the Schrödinger equation:
\begin{equation}
    i\frac{d}{dt}\ket{\psi(t)} = H \ket{\psi(t)},
\end{equation}
whose formal solution gives the unitary time evolution operator (propagator),
\begin{equation}\label{eq:U}
  U(t) = \exp(-i Ht),
\end{equation}
which governs how the quantum state evolves under the action of the Hamiltonian.

Upon Substituting the decomposed form of the Hamiltonian from Eq.~\ref{eq:reducedH} 
into the time evolution operator, we write:
\begin{equation}
  U(t) = \prod_{p<q}^N \exp\left[-i t\left(H_{1}^{pq} + H_{2}^{pq} \right)\right].
\end{equation}
In general, the exponential of a sum of non-commuting operators cannot be factorized 
exactly~\cite{10.1063/1.529425}. 
However, for small time steps, we can approximate the exponential using the 
first-order Trotter–Suzuki decomposition~\cite{Hall2003}.
This approximation neglects the commutator and higher-order terms, and is valid when 
the evolution is broken into sufficiently small intervals.
Applying this, we obtain the approximate form:
\begin{equation}\label{eq:reducedU}
    U(t) \approx \prod_{p<q}^N \exp(-i tH_{1}^{pq})\exp(-i tH_{2}^{pq}).
\end{equation}
This decomposition is significant from the perspective of quantum simulation, as each
exponential term now corresponds to a two-qubit gate acting on a pair of qubits. This makes
the overall time evolution more tractable on near-term quantum devices, where implementing 
multi-qubit gates is challenging and noise-prone.
In the next sections, we simulate a system using quantum circuits
that will perform the collective neutrino oscillations for two-
and three-neutrino system.

\section{Simulation of inversion probability}\label{sec:QuantSim2Neut}
\subsection{For $N = 2$ case : }
For a system consisting of two neutrinos ($N= 2$), we consider the initial state to be composed of one light ($e$) and one heavy ($\mu$) flavour. 
This can be represented as the product state
\begin{equation}\label{eq:prodstate}
  \ket{\psi_0} = \ket{e}\otimes\ket{\mu}, 
\end{equation}
where $\ket{e} (\ket{\mu})$ denotes the electron (muon) flavour state. 
The propagator for this system as in Eq.\eqref{eq:reducedU} is given by 
\begin{equation} \label{eq:N=2}
    U(t) = \exp(-i tH_{1}^{12})\exp(-i tH_{2}^{12}).
\end{equation}
Here, $\exp(-i tH_{1}^{12})$ describes the 
two-flavour neutrino oscillation in vacuum and $\exp(-i tH_{2}^{12})$
describes the neutrino-neutrino interaction. 
The quantum simulation of the vacuum neutrino oscillation term $H_{1}^{12}$ can be performed by encoding the two neutrino flavour states on a single qubit \cite{Arguelles:2019phs}. In Appendix ~\ref{sec:Appen1}, we show the gate structure and quantum simulation of vacuum flavour oscillations. Similarly, the quantum  simulation of the Hamiltonian \eqref{eq:N=2} proceeds by encoding the two neutrinos in two different qubits. The term $H_{2}^{12}$ describing the interaction between two neutrinos can be simulated using two-qubit entangling gates.

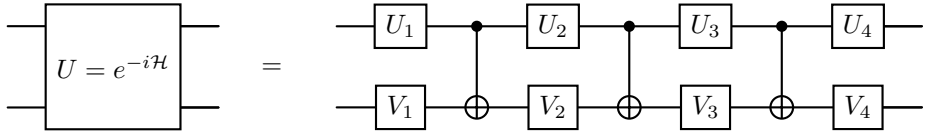
\begin{figure}
    \centering
    \begin{equation*}
    \begin{quantikz}
        \lstick{} & \gate[2]{U = e^{-i\mathcal{H}}} & \qw \\
        \lstick{} &             & \qw
    \end{quantikz}
    \quad = \quad
        \begin{quantikz}
        \lstick{} & \gate{U_1} & \ctrl{1} & \gate{U_2} & \ctrl{1} & \gate{U_3} &
        \ctrl{1} & \gate{U_4} & \qw \\
        \lstick{} & \gate{V_1} & \targ{}  & \gate{V_2} & \targ{}  & \gate{V_3} & 
        \targ{}  & \gate{V_4} & \qw
    \end{quantikz}
\end{equation*}
    \caption{Decomposition of the propagator, $U = e^{-i\mathcal{H}}$ in terms of single 
    qubit unitary gates and CNOT gates. }
    \label{fig:GateArrang}
\end{figure}

To implement the unitary $U(t)$ in Eq.\eqref{eq:N=2} on a quantum circuit, it should be mapped into an appropriate set of gates. An arbitrary two-qubit unitary operation $U \in SU(4)$ can be decomposed into 
a sequence of single-qubit unitary gates and at most three CNOT
gates~\cite{Vidal:2004akd,Vatan:2004nmz}. 
This decomposition is essential for implementing general two-qubit operations
on quantum hardware with a universal gate set.
The theoretical foundations for such decompositions are rigorously developed in 
Ref.~\cite{nielsen2000quantum}, which demonstrate that any two-qubit
gate can be expressed in a canonical form using local operations and an
entangling component. 
In particular, Ref.~\cite{Vidal:2004akd} presents a constructive method for
achieving such a decomposition by identifying a minimal set of CNOT gates and
surrounding single-qubit gates.
The decomposition of a propagator of the form $U = e^{-i\mathcal{H}}$
is shown in Figure~\ref{fig:GateArrang}, where :
\begin{equation}\label{eq:generic-H}
    \mathcal{H} \equiv h_x\sigma_x\otimes\sigma_x + h_y\sigma_y\otimes\sigma_y
    + h_z\sigma_z\otimes\sigma_z,
\end{equation}
where the coefficients $h_x, h_y, h_z \in \mathbb{R}$. On comparing Eq. \eqref{eq:generic-H} with the Hamiltonian, $H_2^{12}$, we obtain :
\begin{equation}
    h_x = h_y = h_z = J_{12}t.
\end{equation} 
The single qubit gates, $U_i$ and $V_i$ in the decomposition \ref{fig:GateArrang} will have the following form \cite{Vidal:2004akd}:
\begin{align}\label{eq:DecompU}
    U_1 &= V_1 = \mathds{1} \nonumber\\
    U_2 &= \frac{i}{\sqrt{2}}(\sigma_x + \sigma_z)
    \exp\big[-i(J_{12}t - \frac{\pi}{4})\sigma_x\big] \nonumber\\
    V_2 &= \exp\big(-i J_{12}t\sigma_z\big) \nonumber\\
    U_3 &= \frac{-i}{\sqrt{2}}\big(\sigma_x + \sigma_z \big) \nonumber\\
    V_3 &= \exp\big(i J_{12}t\sigma_z\big) \nonumber\\
    U_4 &= \frac{\mathds{1} - i\sigma_x}{\sqrt2} \nonumber\\
    V_4 &= \frac{\mathds{1} + i\sigma_x}{\sqrt2}. 
\end{align}
This mapping of the unitary evolution operator into single-qubit and two-qubit entangling gates ensures that the interaction terms in the Hamiltonian are are faithfully represented within the universal decomposition
framework. In Table~\ref{tab:ParamList} we show the value of parameters we considered to simulate the two neutrino collective oscillations. 
\begin{table}[h]
  \caption{Value of parameters considered to simulate the
    two neutrino system}
  \label{tab:ParamList}%
  \begin{tabular}{@{}ll@{}}
    \toprule
    Parameter & Value taken\\
    \midrule
    Mixing angle, $\theta_\nu$ &  0.195 radians~\cite{Hall:2021rbv} \\
    Pair Coupling  angle, $\theta^{12}$   &  $\pi$/6 radians \\
    Squared mass difference, $\Delta m_{12}^{2}$ & 0.0002 eV$^2$\\
    Neutrino energy, $E_{\nu}$ & 0.005 GeV \\
    \botrule
  \end{tabular}
\end{table}

\begin{figure}[hbt!]
  \centering
  \includegraphics[width=1\textwidth]{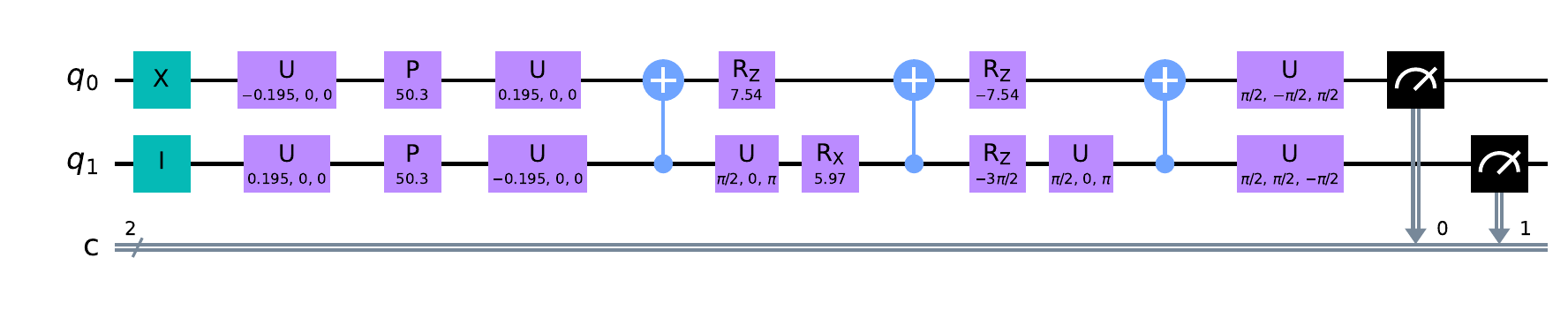}
  \caption{Quantum circuit implementation of a two neutrino system ($N=2$) to simulate the time evolution of collective neutrino effects in dense neutrino gases. Qubits are initialized in the product state of an electron and a muon flavour neutrino (Eq.~\ref{eq:prodstate}). The entangling gates are required to simulate the self-interaction term $H_2^{12}$.The measurements of the encoded qubits gives the desired transition probabilities.}
  \label{fig:circTwoNeutInteraction}
\end{figure}

\begin{figure}[hbt!]
  \centering
  \includegraphics[width=1\textwidth]{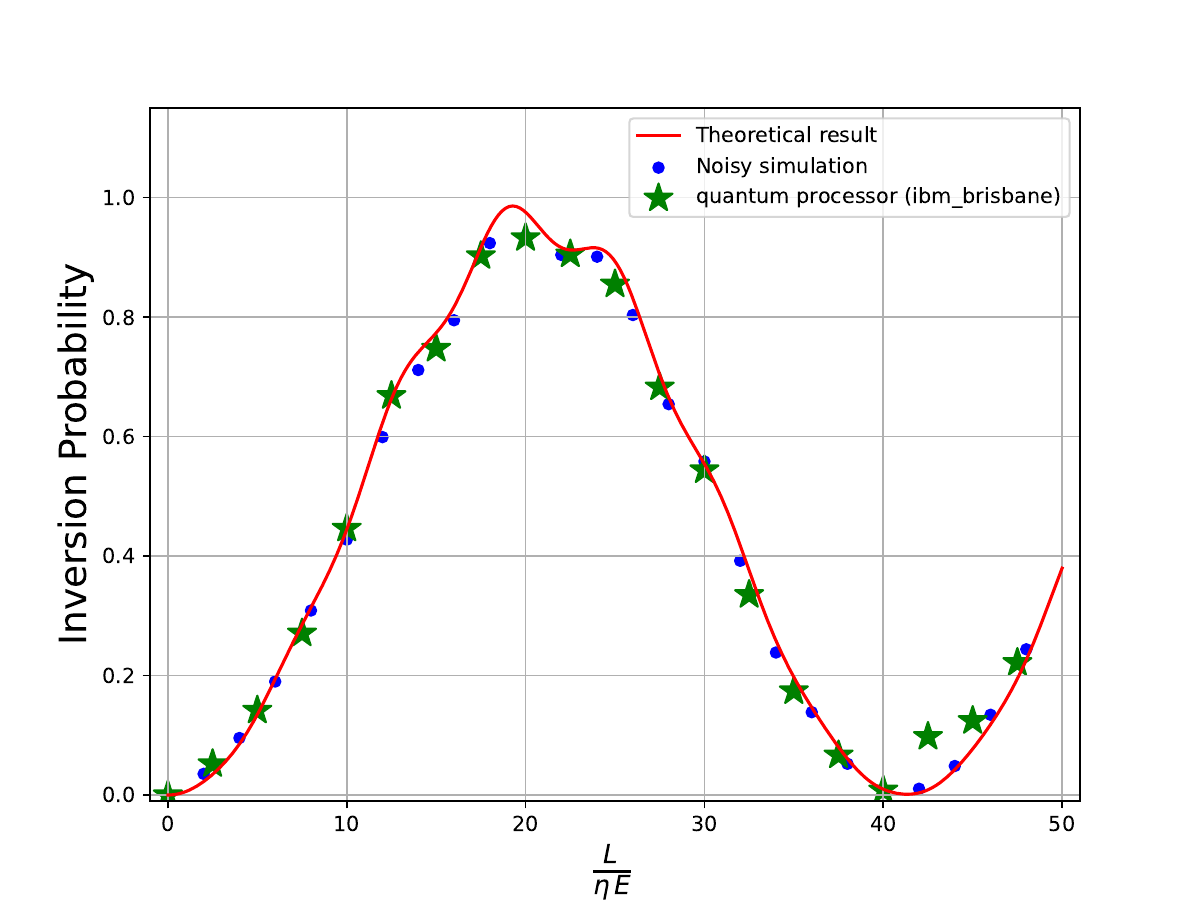}
  \caption{Time evolution of the inversion probability corresponding to the tranisition $\ket{01} \rightarrow \ket{{10}}$.
  The solid red line denotes the theoretical calculation. The data points represents the  results obtained from noisy simulation (\texttt{QiskitAerSimulator}) and IBM QPU (\texttt{ibm\_brisbane}). Each data point is obtained by executing the circuit with 4096 shots.}
  \label{fig:InverProbab}
\end{figure}

Figure~\ref{fig:circTwoNeutInteraction} illustrates the quantum circuit 
used to simulate the time evolution of a two-neutrino system, 
capturing the collective effects present in dense neutrino gases. 
The system is initialized in a product state of one electron- 
and one muon-flavour neutrino, defined by Eq.\ref{eq:prodstate}. 
The qubit $q_1$ is encoded to be in electron-flavour state, i.e. in state 
$\ket{0}$ and $q_2$ represents the muon-flavour state, i.e in state $\ket{1}$, 
which is obtained by applying the Pauli-X gate to $q_0$.  
The first three single-qubit gates positioned implement the vacuum flavour oscillation Hamiltonian $H_1^{12}$ on each of the encoded qubit (see Appendix~\ref{sec:Appen1}). 
The remainder of the circuit, comprising entangling gates such as 
CNOTs along with additional single-qubit rotations, encodes the 
self-interaction Hamiltonian $H_2^{12}$. 
These gates collectively represent the unitary operators $U_i$'s and 
$V_i$'s with $i = (1, 2, 3, 4)$, as described in the decomposition of the 
evolution operator in Eq.\ref{eq:DecompU}. 
Before implementing on a actual quantum processing unit (QPU) the circuit needs to be transpiled to match the hardware's available gate set. In Appendix~\ref{sec:Appen2}, we show the transpiled form of the circuit which is implemented in \texttt{ibm\_brisbane}.

The initial state of the system, $\ket{\psi_0}$, is given in 
Eq.~\ref{eq:prodstate}, i.e., the system starts in the state $\ket{01}$. 
As the system evolves under the Hamiltonian \eqref{eq:reducedH}, the flavour composition
of the neutrinos changes dynamically, leading to flavour transformations. 
This evolution is simulated through the quantum circuit shown in Figure \ref{fig:circTwoNeutInteraction}.  In this circuit, the encoded qubits undergo coherent oscillations though a sequence of unitary gates designed to mimick the neutrino flavour oscillations. 
In particular, if the system transitions from the state $\ket{01}$ to the state $\ket{10}$, it is 
referred as inversion. 
The probability of such an inversion is given by:
\begin{equation}
    P_{{inv}} = \big| \braket{10 | \psi(t)} \big|^2,
\end{equation}
where $\ket{\psi(t)}$ denotes the quantum state of the system at time $t$.

In Figure \ref{fig:InverProbab}, we show the quantum simulation of the time evolution of inversion probability  and compare it with theoretical results.
The solid red line denotes the theoretical calculation. The data points represents the 
results obtained from noisy simulation (\texttt{QiskitAerSimulator}) and IBM QPU 
(\texttt{ibm\_brisbane}). All the data points shows good agreement with theory. 

\subsection{For $N = 3$ Case : }
\begin{figure}[b]
    \centering
    \includegraphics[width=1\linewidth]{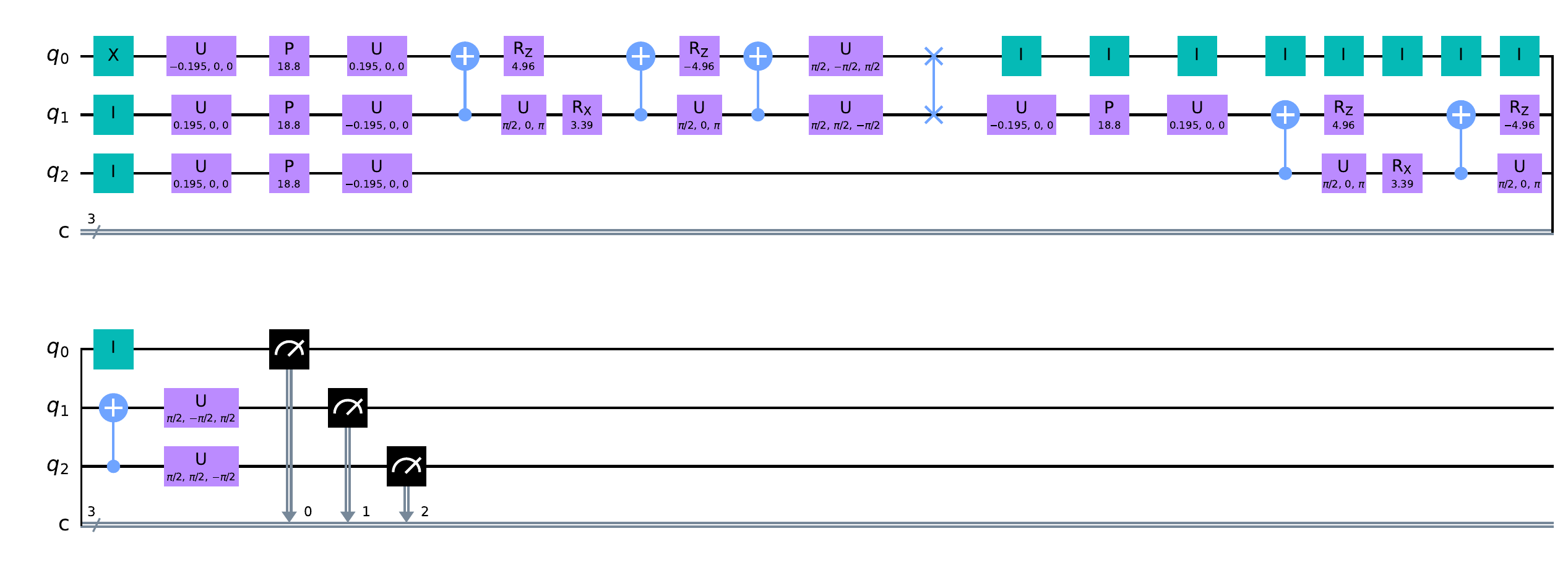}
    \caption{Circuit representation of the three-neutrino system to simulate 
    the time evolution of collective neutrino effects.}
    \label{fig:circThreeNeutInt}
\end{figure}

In the case of three-neutrino system, the Hamiltonian includes pairwise interaction between all the three neutrinos. The encoding of the neutrino flavour states onto three qubit state is done in a similar manner:
\begin{equation} \label{eq:prodstate3Neut}
    \ket{\psi_0} = \ket{e} \otimes \ket{e} \otimes \ket{\mu}. 
\end{equation}
The quantum circuit now has three interacting qubits which encode the evolution of neutrinos with three interacting Hamiltonian terms: $H_2^{12}$, $H_2^{23}$ and $H_2^{13}$. 
%For the $e-e$ pair though, 
The quantum circuit representing the three-neutrino system to simulate the time 
evolution of inversion probability is shown in Figure~\ref{fig:circThreeNeutInt}.
Qubits $q_1$ and $q_2$ encode the $\ket{\nu_e}$ and qubit $q_0$ is 
encoded to be in $\ket{\nu_\mu}$ state. Now, in addition to the unitaries for the pairwise interaction simulation there, the circuit involves a SWAP gate to exchange the positions of $\ket{\nu_\mu}$ and  $\ket{\nu_e}$. 

\begin{figure}[hbt!]
    \centering
    \includegraphics[width=1\linewidth]{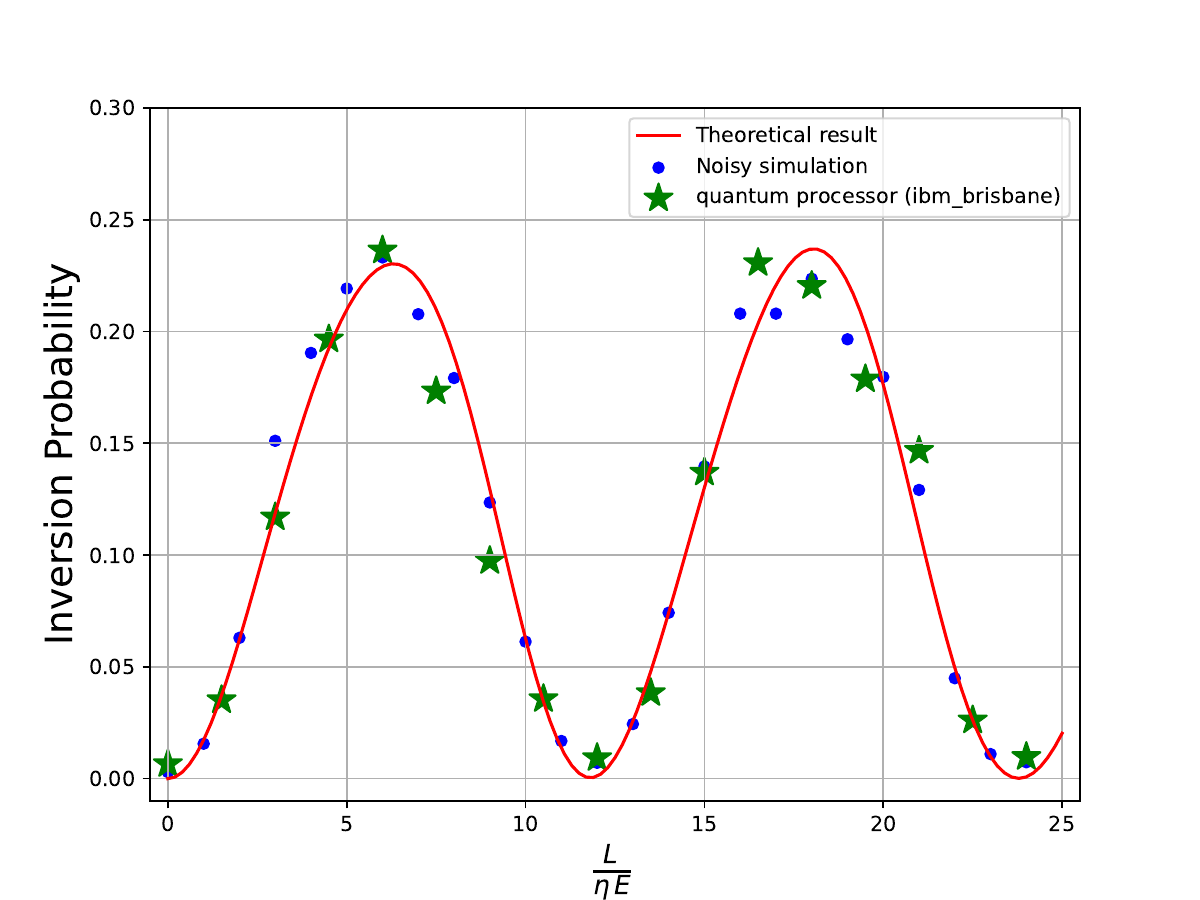}
    \caption{Time evolution of the inversion probability corresponding to the tranisition $\ket{001} \rightarrow \ket{{100}}$.
  The solid red line denotes the theoretical calculation. The data points represents the  results obtained from noisy simulation (\texttt{QiskitAerSimulator}) and IBM QPU (\texttt{ibm\_brisbane}).  Each data point is obtained by executing the circuit with 4096 shots.}
    \label{fig:InverProbab3Neut}
\end{figure}

The probability of inversion of the state $\ket{001}$, in this case is given by:
\begin{equation}
    P_{{inv}} = \big| \braket{100 | \psi(t)} \big|^2,
\end{equation}
where $\ket{\psi(t)}$ denotes the quantum state of the three-neutrino 
system at time $t$. In Table~\ref{tab:ParamList3Neut} we show the value of parameters considered
to simulate the three-neutrino system. 
\begin{table}[h]
  \caption{Value of parameters considered to simulate the
    three-neutrino system}
  \label{tab:ParamList3Neut}%
  \begin{tabular}{@{}ll@{}}
    \toprule
    Parameter & Value taken\\
    \midrule
    Mixing angle, $\theta_\nu$ &  0.195 radians~\cite{Hall:2021rbv} \\
    Pair Coupling  angles & \\
    \hspace{1cm}$\theta^{12}$   &  0 radians \\
    \hspace{1cm}$\theta^{13}$   &  $\pi$/6 radians \\
    \hspace{1cm}$\theta^{23}$   &  $\pi$/6 radians \\
    Squared mass difference, $\Delta m_{12}^{2}$ & 0.0002 eV$^2$\\
    Neutrino energy, $E_{\nu}$ & 0.005 GeV \\
    \botrule
  \end{tabular}
\end{table}
In Figure~\ref{fig:InverProbab3Neut} we show the time evolution of the inversion 
probability for these set of parameters. 
The solid red line denotes the theoretical calculation. Data points obtained from
Noisy simulator (\texttt{QiskitAerSimulator}) and IBM processor (\texttt{ibm\_brisbane}) 
are shown by blue dots and green stars respectively. All data points shows a 
good agreement with the theoretical results.

\section{Quantum circuit for calculating entanglement}\label{sec:Entang}
In this section, we study the entanglement between the two neutrinos by calculating the concurrence of the neutrino state. 
For a bi-partite system, the entanglement of a pure state is defined 
as the entropy of either of the two subsystems~\cite{Bennett:1995tk} : 
\begin{equation}
    E(\psi) = -\mbox{Tr}(\rho \log_2 \rho), 
\end{equation}
where $\rho$ is the density matrix of one subsystem, obtained by 
tracing over the degrees of freedom of other subsystem.
It can also be written in terms of concurrence as~\cite{Hill:1997pfa} :
\begin{equation}
    E(\psi) = {E(C(\psi))}.
\end{equation}
Concurrence measures the absolute value of fidelity of
an arbitrary state onto the spin-flipped state of itself ~\cite{Wootters:1997id}.
It returns the value in the range $[0, 1]$, with 0 being the system in the
product state, i.e. with no entanglement and 1 implies the maximum 
entanglement. 
Spin-flipped state of an arbitrary state, $\ket{\psi}$ is obtained by
applying the spin-flip operation ($\sigma_y$) to its complex conjugate :
\begin{equation}\label{eq:spinflip}
  \ket{\tilde{\psi}} = \sigma_y\ket{\psi^{*}}.
\end{equation}
To get the spin-flipped state of a product state of $N$ qubits, 
one must apply the flipping operation to all individual states. 
The concurrence of the state $\ket{\psi}$ is then written as :
\begin{equation}\label{eq:concurr}
  C(\psi) = \big|\braket{\psi|\tilde{\psi}} \big|.
\end{equation}
To implement its formulation on a quantum processor, we first encode the 
state $\ket{\psi}$ onto the sets of qubits, like we did in previous section. 
To obtain the spin-flipped state from $\ket{\psi}$ we
apply the operator $\sigma_y$ to each of the qubits. 
Several parameters, like those for Phase-gate, $R_z$-gate along with the
coefficients of the Hamiltonian, $h_x, h_y, h_z$ are reversed for the
state $\ket{\psi^*}$.
We also require an ancilla qubit in the circuit whose measurements give us the concurrence. Thus, the total qubits 
required to calculate the concurrence for $N$-neutrino
system is $2N+1$. 
The probability of survival of the ancilla qubit, which is
initially prepared in state $\ket{0}$ and is entangled
to the neutrino states as : 
\begin{equation}
  P(0) = \frac{1}{2}\Big(1 + \big|\braket{\psi_0|\tilde{\psi_0}}
  \big|^2 \Big).
\end{equation}
The concurrence of the state $\ket{\psi}$ is then calculated by the formula :
\begin{equation}\label{eq:concurranc}
    C(\psi) = \sqrt{2P(0) - 1}.
\end{equation}
The method of calculating the absolute value of inner product of two
state using an ancilla qubit is shown in Appendix~\ref{sec:Appen2}.

\begin{figure}[hbt!]
  \centering
  \includegraphics[width=1\textwidth]{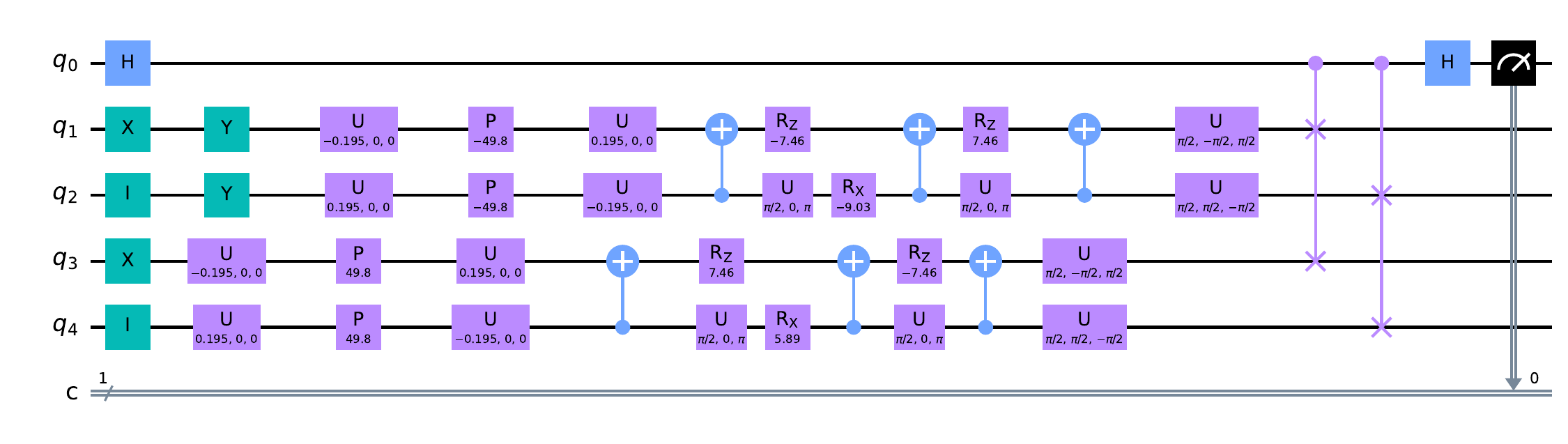}
  \caption{Circuit representation of the calculation of concurrence of
    the entangled states in a two-neutrino system.
    Here, the qubits $q_3$ and $q_4$ forms the product state
    $\ket{\psi_0}$ and $q_1$ and $q_2$ form its spin-flipped state.
    $q_0$ represents the ancilla qubit.}
  \label{fig:ConcurrenceCirc}
\end{figure}

\begin{figure}[hbt!]
  \centering
  \includegraphics[width=1\textwidth]{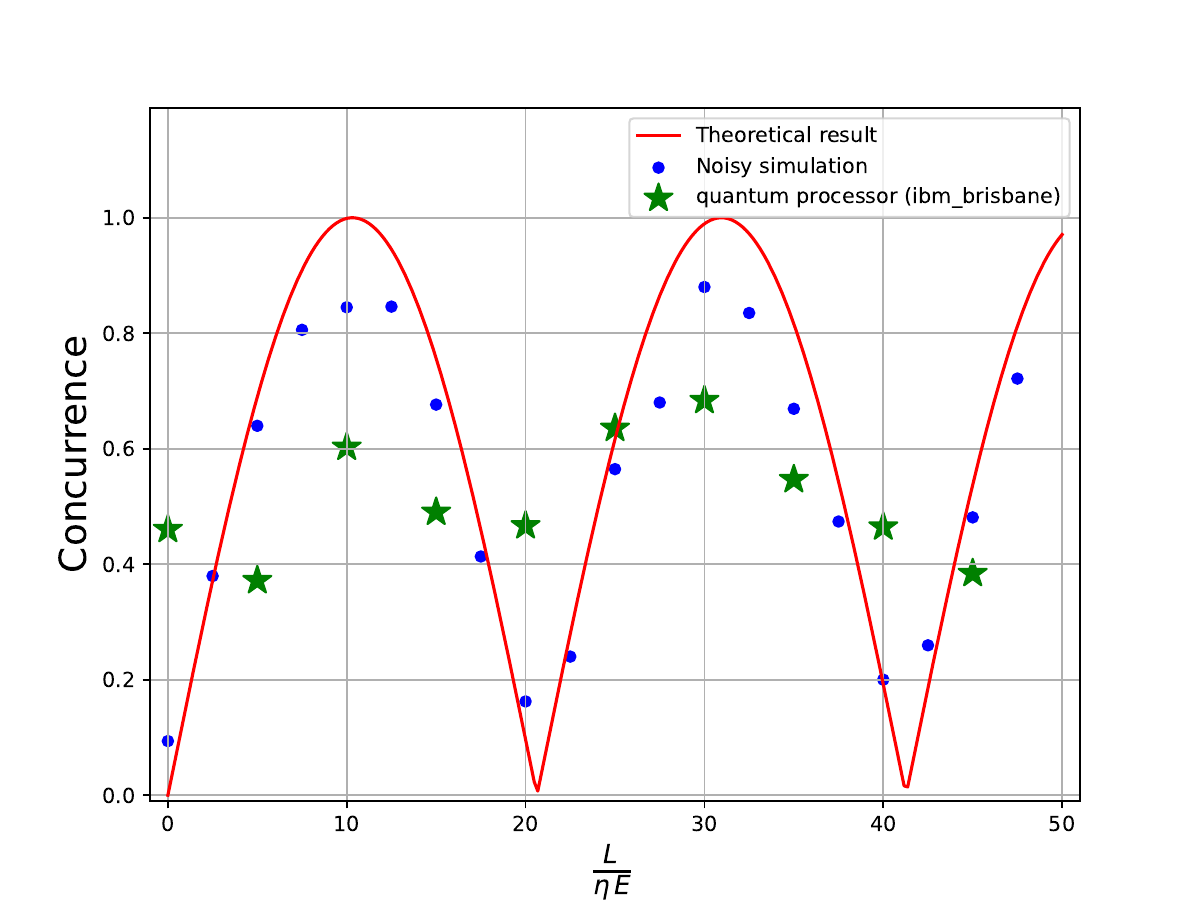}
  \caption{Time evolution of the concurrence, given by Eq.~\ref{eq:concurranc}.
  The solid line represents the theoretical calculation. 
  Blue dots and green stars represent the data points from the simulation
  on noisy simulator (\texttt{QiskitAerSimulator}) and quantum processor (\texttt{ibm\_brisbane}).}
  \label{fig:Concurrence}
\end{figure}

After encoding the initial product state of two-neutrino
system on two qubits, two more qubits were utilised to encode the
spin-flipped state.
The quantum circuit for the calculation of concurrence of
the entangled states in a two-neutrino system is shown in
Figure~\ref{fig:ConcurrenceCirc}.
Here, the qubits $q_3$ and $q_4$ forms the product state
$\ket{\psi_0}$ and $q_1$ and $q_2$ form its spin-flipped state
while $q_0$ represents the ancilla qubit.
Figure~\ref{fig:Concurrence} shows the time evolution of the concurrence (~\ref{eq:concurr}). 
The solid line represents the theoretical calculation. 
Blue dots and green stars represent the data points from the simulation
on noisy simulator (\texttt{QiskitAerSimulator}) and quantum processor (\texttt{ibm\_brisbane}).
From the figure, we can see that initially the value of concurrence is zero, which is 
expected as the system initially is in the product state of individual neutrino
flavour state. 
As the system propagates in time, the state becomes entangled due to flavour 
mixing and swapping between the neutrinos. 
This entanglement progresses periodically with time. 
If compare this with inversion probability (Figure \ref{fig:InverProbab} ) for two-neutrino system, 
we notice that whenever the probability of inversion is minimum and maximum, 
the concurrence is zero, i.e. we have no entanglement. 
The maximum entanglement occurs exactly at times when the probability of 
inversion is half of the maximum value. 
Results from the noisy simulation shows better match with the theoretical 
results. 
The effects of noise can be seen in the form of small dip in the
values when compared with the theoretical results.
These noises are the result of decoherence and gate errors that
degrade the quantum state's purity.

\section{Summary and Conclusions}\label{sec:Summ}
In this work, we simulate neutrino flavour state in 
a core-collapse supernova-like environment 
on a quantum processor. 
In such dense neutrino gases, the extremely high neutrino density 
leads to significant self-interactions, resulting in flavour 
conversion phenomena.
In addition to vacuum and matter-induced mixing, these environments
introduce an extra term in the Hamiltonian due to neutrino-neutrino 
forward scattering via Z-boson exchange.
To simplify the problem, we adopt common assumptions such as spatial 
homogeneity and isotropy, even though real supernova environments 
are neither.
We also neglect the MSW effect in regions near the neutrino sphere,
where neutrino self-interaction dominates. 
As illustrative examples, we consider two- and three-neutrino 
systems to study the propagation of initial states composed of 
light ($e$) and heavy ($\mu$) flavours.

We have encoded the time evolution of neutrino states on a 
quantum processor and evaluated the probability of state inversion,
comparing the results with theoretical predictions. 
To quantify quantum correlations, we calculated the concurrence
between qubits representing neutrino flavours and validated the
outcomes against theoretical expectations.

This study demonstrates a foundational approach to simulate the systems 
that mimic dense neutrino gases on quantum hardware, specifically 
for $N=2$ and $N=3$ number of neutrinos.
The methodology can be extended to many-neutrino systems, where 
classical computation becomes infeasible due to exponential complexity. 
Such investigations not only advance our understanding of neutrino
physics—including their flavour evolution and interaction dynamics in 
astrophysical environments—but also supports the case of development of scalable 
quantum processors capable of simulating systems with all-to-all interactions.
The convergence of astro-particle physics and quantum computing holds 
promise for building deeper phenomenological models and driving innovation
in both fields.

\section*{Acknowledgements}
We acknowledge the use of Qiskit AerSimulator and \texttt{ibm\_brisbane} in this study. The views expressed are those of the authors and do not reflect the official policy or position of IBM or the IBM Quantum team. 
%\clearpage
\begin{appendices}

  \section{Two flavour neutrino oscillations using single qubit on a
    quantum computer}\label{sec:Appen1}
  For two flavour oscillations, we have considered the case of 
  survival and disappearance probability of $\nu_e$.
  The theory of development is given in Ref~\cite{Thomson:2013zua}.
  The probability of disappearance of an $\nu_e$ during two flavour 
  oscillations comes out to be :
  \begin{equation} 
    P(\nu_{e} \rightarrow \nu_{\mu}) = \sin^{2}(2\theta_\nu)\sin^{2}
    \Big(\frac{(m^{2}_{1} - m^{2}_{2})L}{4E_\nu}\Big),
  \end{equation}
  where, $\theta_\nu$ is the mixing angle, $m_{1}$ and $m_{2}$ are 
  the masses of neutrino mass eigenstates corresponding to $\nu_{1}$
  and $\nu_{2}$ respectively, $L$ is the distance travelled by $\nu_e$
  with energy $E_\nu$.
  We can express the disappearance probability in the units of length 
  and energy scales which are used in practice as :
  \begin{equation} \label{eq:ProbabOsc1}
    P(\nu_{e} \rightarrow \nu_{\mu}) = \sin^{2}(2\theta_\nu)\sin^{2}
    \Big(1.27\frac{\Delta m^{2}[eV^{2}]L[km]}{E[GeV]}\Big).
  \end{equation}

  To simulate the two-flavour neutrino oscillations in vacuum 
  on a quantum processor,  
  we encode the two flavour eigenstates $\ket{\nu_e}$ and $\ket{\nu_\mu}$
  in the two states of a qubit, $\ket{0}$ and $\ket{1}$ respectively.
  The transformation of flavour basis into mass basis can be studied 
  using a $2\times2$ unitary matrix called PMNS matrix. 
  The detailed analysis can be found in Ref~\cite{Arguelles:2019phs}.

\begin{figure}[hbt!]
    \centering
    \includegraphics{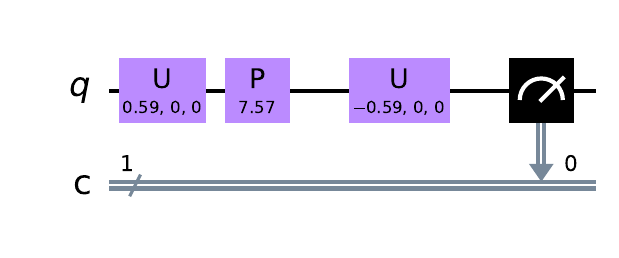}
    \caption{Circuit diagram to calculate the probabilities of
      two flavour neutrino oscillations.}
    \label{fig:CircTwoFlav}
  \end{figure}

\begin{figure}[hbt!]
    \centering
    \includegraphics[width=0.48 \textwidth]{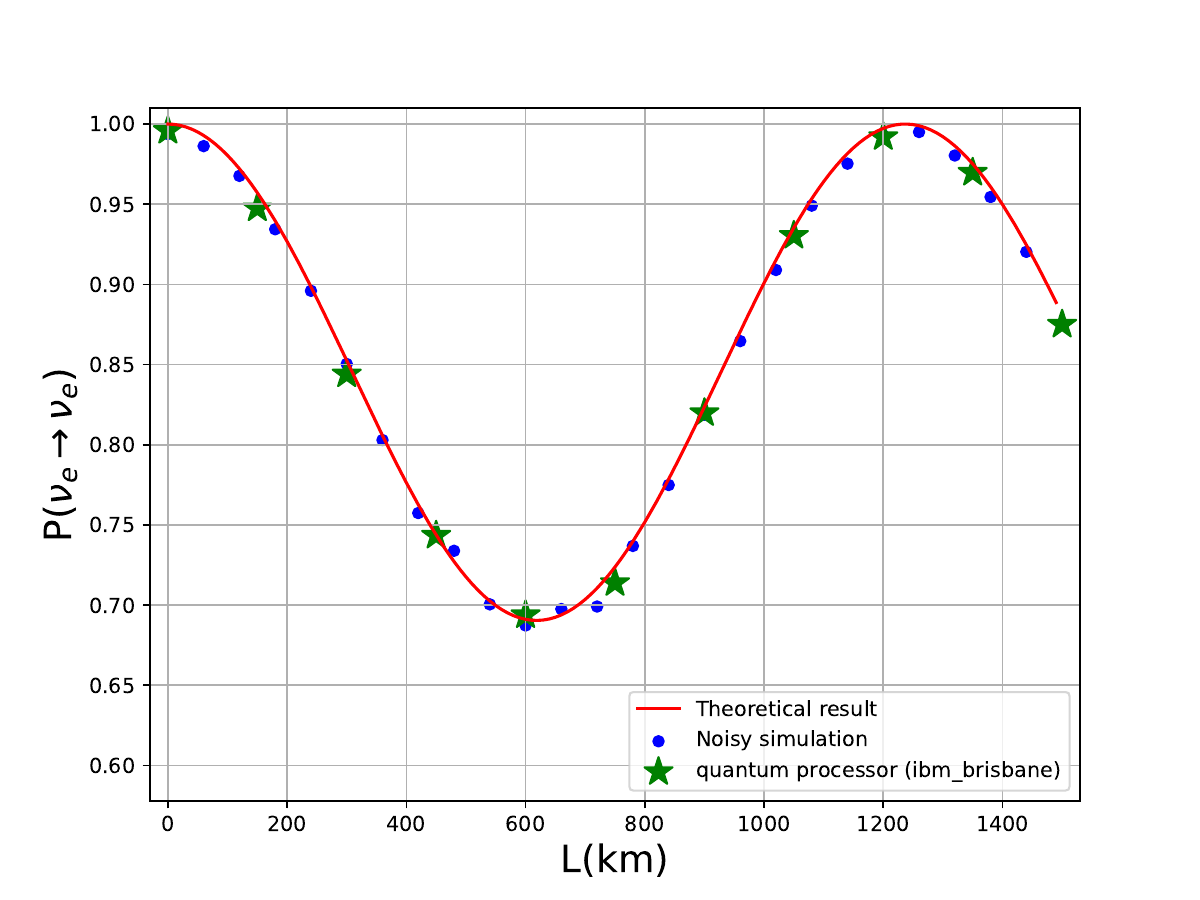}
    \includegraphics[width=0.48 \textwidth]{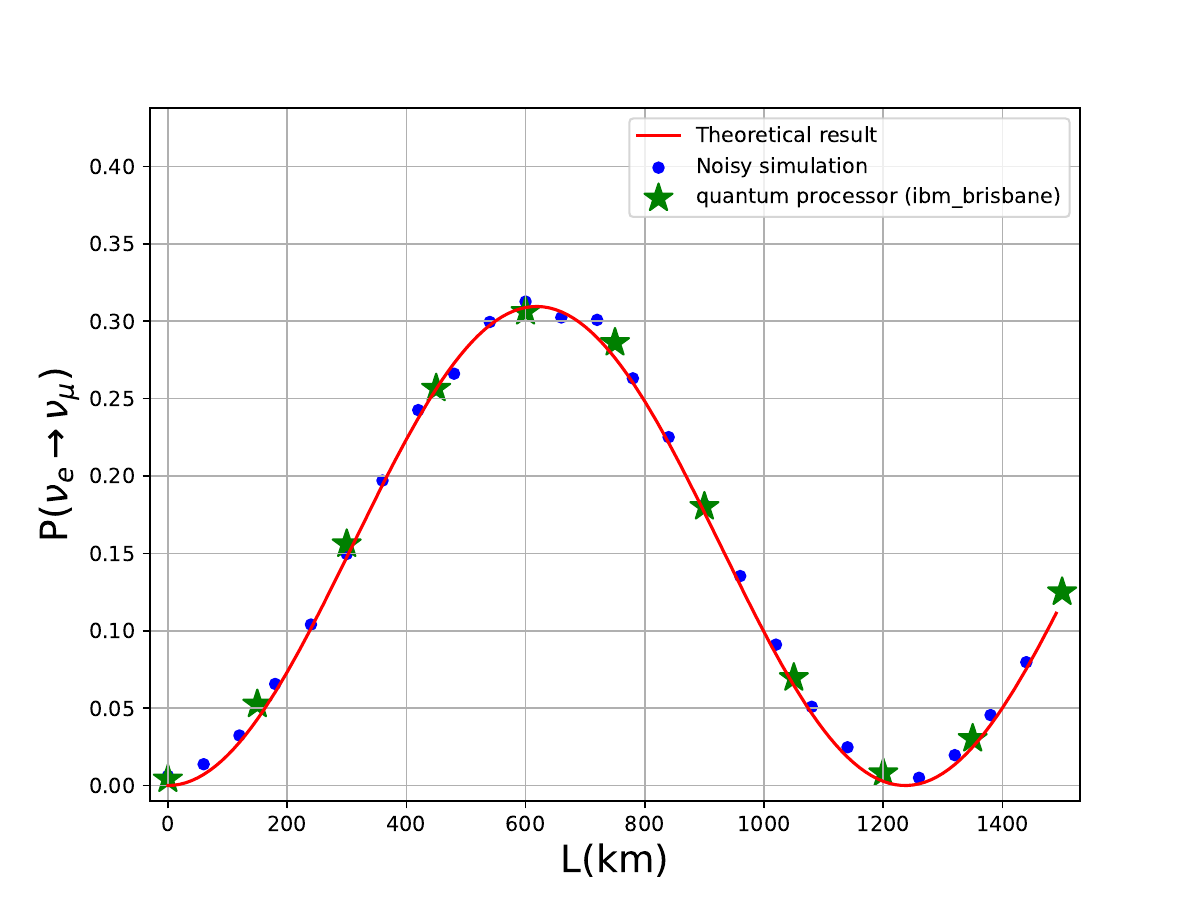}
    \caption{Time evolution of 
  survival probability (left panel) and disappearance probability 
  (right panel) of an $\nu_e$. 
  The solid red line denotes the theoretical result. Noiseless and 
  Noisy simulation results are denoted by blue dot and green stars 
  respectively.}
    \label{fig:OscProbTwoFlav}
  \end{figure}

  Figure~\ref{fig:CircTwoFlav} shows the circuit representation
  of a single qubit system which simulates the two-flavour vacuum 
  oscillations of a neutrino which is initially being 
  an electron-type (state $\ket{0}$ of the qubit). 
  The two unitary gates at each ends of the qubit line are for 
  the basis transformations.
  The phase gate at the middle governs the time evolution of 
  mass eigenstates. 
  The value of parameters used to simulate this system are given in 
  Table~\ref{tab:ParamListFlavOsc}.
  \begin{table}[h]
  \caption{Value of parameters considered to simulate the
    two flavour neutrino oscillation using single qubit}
  \label{tab:ParamListFlavOsc}%
  \begin{tabular}{@{}ll@{}}
    \toprule
    Parameter & Value taken\\
    \midrule
    Mixing angle, $\theta_\nu$ &  0.295 radians \\
    Squared mass difference, $\Delta m_{12}^{2}$ & 0.0002 eV$^2$\\
    Neutrino energy, $E_{\nu}$ & 0.005 GeV \\
    \botrule
  \end{tabular}
\end{table}  
  Figure~\ref{fig:OscProbTwoFlav} shows the time evolution of 
  survival probability (left panel) and disappearance probability 
  (right panel) of an $\nu_e$. 
  The solid red line denotes the theoretical result. Noiseless and 
  Noisy simulation results are denoted by blue dot and green stars 
  respectively. 
  The maximum flavour mixing probability is given by $\sin^{2}2\theta_\nu$
  in Eq.~\ref{eq:ProbabOsc1}, which comes out to be
  0.3095 for the given parameters. 
  
  \section{Qiskit Simulator : Layout and Transpilation} \label{sec:Appen2}
  The \texttt{ibm\_brisbane} backend features a heavy-hexagonal qubit layout, 
  designed to minimize crosstalk and reduce gate errors by ensuring limited nearest-
  neighbour connectivity. 
  This layout is a characteristic feature of IBM’s Falcon and Eagle quantum processors.
  In particular, Eagle-type processors implement scalable heavy-hex connectivity
  with qubits arranged in a repeating hexagonal tiling pattern across layers, 
  enabling improved coherence and fidelity.   The \texttt{ibm\_brisbane} QPU is one of the IBM Eagle processor having 127 superconducting qubits.
  The native gate set available in this processor include ECR (echoed cross-resonance, two qubit gate), RZ (Single qubit $Z$ rotation), $\sqrt{X}$ (single qubit $\sqrt{NOT}$ gate), X (single qubit NOT gate) and ID (single qubit Identity) gates.

\begin{figure}
      \centering
      \includegraphics[width=1\linewidth]{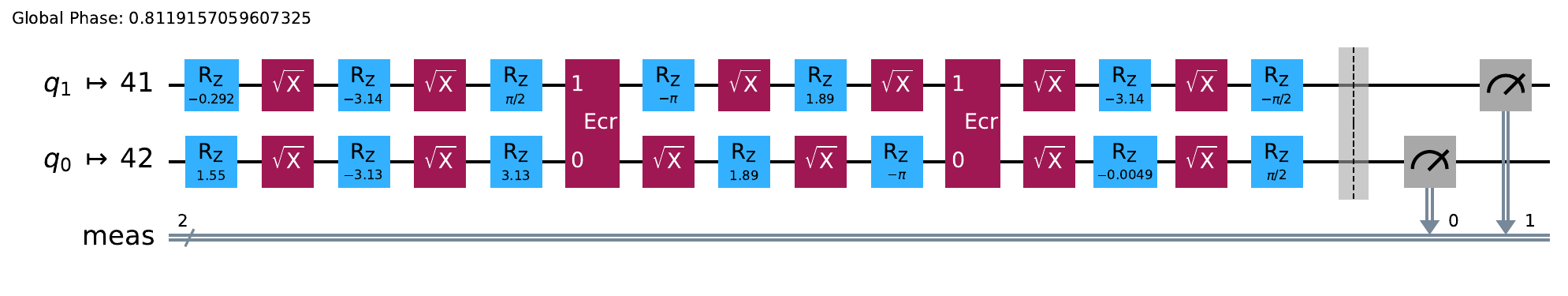}
      \caption{Transpiled circuit from the \texttt{ibm\_brisbane} simulator
      used in the simulation of calculation of inversion probability 
      of a two-neutrino system, where the initial state $\ket{01}$, as 
      given by Eq.~\ref{eq:prodstate}.}
      \label{fig:simulatorcirc}
  \end{figure}

  The Qiskit \texttt{AerSimulator} allows high-performance classical simulation of 
  quantum circuits, either in an ideal (noise-free) setting or by incorporating
  realistic noise models.  \texttt{AerSimulator} assumes idealized full connectivity unless a specific coupling map is defined. This makes it useful for algorithm development and benchmarking without hardware constraints. 
  We perform noisy simulations using \texttt{AerSimulator} and the 
  noise model derived from \texttt{ibm\_brisbane} backend.
  By importing its noise model into the \texttt{AerSimulator}, we closely mimic
  hardware-level imperfections in a controlled environment. 
  This approach allows us to compare the theoretical (ideal), noisy simulated,
  and hardware-executed circuit results. 
  The noisy simulations provide insight into the effects of quantum noise and 
  help validate the robustness of our algorithm before deployment on the real device.
  Figure~\ref{fig:simulatorcirc} shows the transpiled circuit implemented in  \texttt{ibm
  \_brisbane} which is used in the simulation of calculation of inversion probability 
  of two-neutrino system (Figure \ref{fig:circTwoNeutInteraction}).

  \section{Calculating the absolute value of inner product of two arbitratry 
  quantum states using ancilla qubit} \label{sec:Appen3}      
  Suppose initially we have two quantum states, $\ket{\psi}$ and $\ket{\phi}$. 
  Calculation of the absolute value of their inner product, 
  $\big|\braket{\psi|\tilde{\psi}} \big|$ on a quantum processor can be done
  using following steps :
  \begin{description}
      \item[\textbf{Step 1 :}] Introduce an ancilla in state $\ket{0}_{anc}$ such that 
      the product state of the system is :
      \begin{equation} 
          \ket{\chi}_i = \ket{0}_{anc} \otimes \ket{\psi} \otimes \ket{\phi}.
      \end{equation}
      \item[\textbf{Step 2 :}] Apply Hadamard gate to the ancilla to get the following 
      state :
      \begin{equation}
          H_{anc} \ket{\chi}_i = \frac{1}{\sqrt{2}}\Big(\ket{0}_{anc}+\ket{1}_
          {anc}\Big)\otimes \ket{\psi} \otimes \ket{\phi}.
      \end{equation}
      \item[\textbf{Step 3 :}] Apply Controlled-SWAP (Fredkin gate), with ancilla 
      being the control qubit, which will perform the swapping 
      between states $\ket{\psi}$ and $\ket{\phi}$ only if the ancilla is in state $\ket{1}$.  
      \begin{align}
          CSWAP_{anc,\ket{\psi},\ket\phi}H_{anc}\ket{\chi}_i = 
          \frac{1}{\sqrt2}\Big(&\ket{0}_{anc}\otimes\ket{\psi}\otimes\ket{\phi} 
          + \ket{1}_{anc}\otimes\ket{\phi}\otimes\ket{\psi} \Big).
      \end{align}
      \item[\textbf{Step 4 :}] Apply another $H_{anc}$ on the ancilla to get : 
      \begin{align}
          \ket{\chi}_f &= H_{anc}CSWAP_{anc,\ket{\psi},\ket{\phi}}H_{anc}
          \ket{\chi}_i \nonumber \\
          &= \frac{1}{2}\Bigg[ \ket{0}_{anc}
          \Big(\ket{\psi}\ket{\phi} + \ket{\phi}\ket{\psi}\Big) 
          + \ket{1}_{anc}\Big(\ket{\psi}\ket{\phi}
          - \ket{\phi}\ket{\psi}\Big) \Bigg].
      \end{align}
  \end{description}
  Survival probability of ancilla state is given by :
  \begin{align}
    P(0) &= \Bigg| \frac{1}{2}\Big( \ket{\psi}\ket{\phi}
    + \ket{\phi}\ket{\psi}\Big)\Bigg|^2    
    = \frac{1}{2}\Big( 1 + \big|\braket{\psi|\phi} \big| ^2\Big)
  \end{align}
  If we take state $\ket{\phi}$ to be the spin-flipped state of $\ket{\psi}$
  then the concurrence from Eq.\ref{eq:concurr} is : 
  \begin{equation}
      C = \big|\braket{\psi|\tilde{\psi}} \big| = \sqrt{2P(0) - 1}
  \end{equation}

\end{appendices}

%\clearpage
%\bibliographystyle{sn-article}
\bibliography{sn-article}% common bib file

\end{document}